\documentclass[12pt,preprint]{aastex}

\shorttitle{A hard and variable X-ray emission from HD\,157832}
\shortauthors{R. Lopes de Oliveira \& C. Motch}

\begin{document}

\title{A hard and variable X-ray emission from the massive emission line star HD\,157832}

\author{R. Lopes de Oliveira}
\affil{Universidade de S\~ao Paulo, Instituto de F\'isica de S\~ao Carlos, Caixa Postal 369, 13560-970, S\~ao Carlos, SP, Brazil}

\and

\author{C. Motch}
\affil{Universit\'e de Strasbourg, CNRS UMR 7550, Observatoire Astronomique, 11 rue de l'Universit\'e, F67000 Strasbourg, France}

\begin{abstract}
We report the discovery of a hard-thermal (T\,$\sim$\,130\,MK) and variable X-ray emission from the Be star HD\,157832, a new member of the puzzling class of $\gamma$-Cas-like Be/X-ray systems. Recent optical spectroscopy reveals the presence of a large/dense circumstellar disc seen at intermediate/high inclination. With a B1.5V spectral type, HD\,157832 is the coolest $\gamma$-Cas analog known. In addition, its non detection in the ROSAT all-sky survey shows that its average soft X-ray luminosity varied by a factor larger than $\sim$ 3 over a time interval of 14\,yr. These two remarkable features, ``low'' effective temperature and likely high X-ray variability turn HD\,157832 into a promising object for understanding the origin of the unusually high temperature X-ray emission in these systems.
  
\end{abstract}


\section{Introduction}

Massive stars emit thin thermal soft X-rays characterized by a plasma temperature of $\sim$ 6\,MK ($\equiv$ $kT$\,$\sim$\,0.5\,keV), most likely due to the thermalization of part of the fast stellar wind in shocks \citep[e.g.,][]{Gudel09}. 
Hotter X-ray temperatures are observed in massive colliding wind binaries \citep[$\sim$ 23 MK; e.g.,][]{Becker06} and in the case of magnetically channeled wind shocks \citep[up to $\sim$ 45 MK; e.g.,][]{Donati09}. 
However, the X-ray emission of the Be star $\gamma$ Cassiopeiae (\object{$\gamma$-Cas})
is at variance with that of other massive stars: it is dominated by a very hot component (T\,$\sim$\,140--165\,MK), with minor contributions of 2 or 3 colder plasmas having temperatures in the range from 1 MK to 35 MK \citep[][]{Smith04,Lopes10}. Moreover, \object{$\gamma$-Cas} displays a moderately high X-ray luminosity (10$^{32-33}$\,erg\,s$^{-1}$;  0.2--12\,keV) variable on time scales ranging from a few seconds to several weeks, in contrast with the nearly stable flux and one order of magnitude lower luminosity usually observed from massive stars of similar spectral types. 

In spite of being one of the most frequently observed stars in X-rays, the nature of the high energy emission of \object{$\gamma$-Cas} remains a puzzle. On one hand, its X-ray spectrum resembles that emitted by many cataclysmic variables, supporting the idea that accretion onto a white dwarf is responsible for the high energy emission. The thin thermal nature of the X-ray spectrum of \object{$\gamma$-Cas} is unlike the non thermal power-law energy distribution typical of all known accreting neutron stars Be/X-ray binaries and thus hints at a different kind of compact object. On the other hand, some indirect evidence of magnetic activity and simultaneous variations of the X-ray, optical, and UV emission support the idea that its X-rays are produced by magnetic reconnection at the interface between the photosphere and the inner part of the disc \citep[][]{Smith02}.

\object{$\gamma$-Cas} remained a unique object for about 20 years. The recent discovery of at least six new $\gamma$-Cas analogs in X-ray surveys \citep{M07,Lopes06,Lopes07,SB06,Safi07,Rakowski06} now allows us to find additional clues to the X-ray emission mechanism.
We report here on the discovery of a new and peculiar $\gamma$-Cas analog, the Be star \object{HD\,157832}.

\begin{figure*}
\centerline{
\includegraphics[scale=.7,angle=-90]{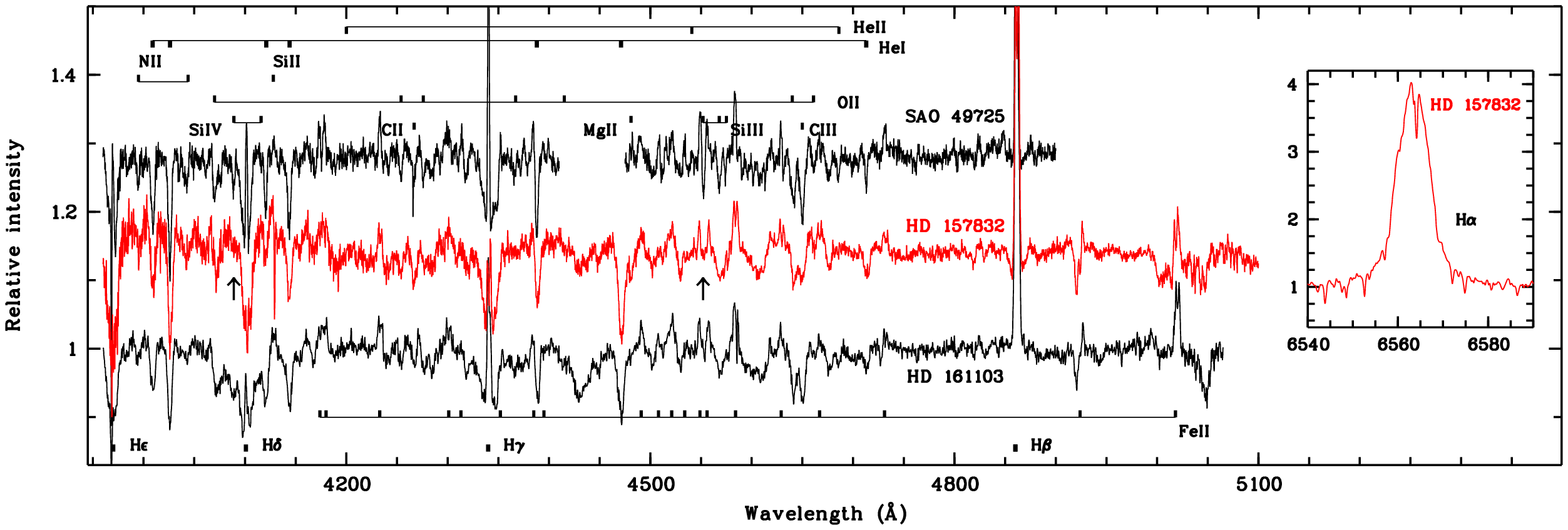}
}
\caption{Optical spectrum of HD\,157832. The spectra of the $\gamma$-Cas-like stars HD\,161103 and SAO\,49725 are shown for comparison \citep[after][]{Lopes06}.\label{fig1}}
\end{figure*}

\section{X-ray and Optical Observations}

\object{HD\,157832} was serendipitously observed by the EPIC cameras on-board the XMM-{\it Newton} observatory during 21.9 ks on 2008 September 5 (ObsID 0551020101). Its X-ray counterpart is the bright source 2XMMi J172754.8-470134. All EPIC cameras were operated in the {\it full frame} mode using the {\it medium} filter which efficiently rejected optical light from the target of the observation GJ\,674 and from \object{HD\,157832}.
The background flare-free good time interval is 17.6 ks for the PN camera and 21.7 ks for the MOS1 and MOS2 cameras. The data were reprocessed and reduced with Science Analysis System (SAS) software v10. 

We obtained optical spectra of \object{HD\,157832} on 2010 November 9 with the {\it Coud\'e} spectrograph mounted on the Brazilian 1.6-m Perkin-Elmer telescope at the {\it Pico dos Dias} Observatory.\footnote{administered by the National Astrophysics Laboratory (LNA; Brazil)} Two position angles of the 600 l/mm grating allowed us to cover the blue (3960-5100 \AA; two exposures) and the H$\alpha$ (5900-7000 \AA; one spectrum) regions, with a dispersion of $\sim$ 0.25 \AA/pixel. Data reduction was done with the MIDAS/ESO software following standard procedures.

\section{Results}

\subsection{Optical properties}

\object{HD\,157832} is a bright star (V $\sim$ 6.6 mag) which has so far barely attracted attention. A rather large range of spectral types are reported in the literature: B0Vne \citep{Buscombe95}, B3IVe \citep{Jaschek92}, and B5Vnne \citep{Thackeray73}. 
Our optical spectrum (see Fig. \ref{fig1}) shows, indeed, somewhat ambiguous signatures. The lack of Si {\small IV} $\lambda$4089 line indicates a spectral type later than B1V, or later than B2V from the weak Si {\small III} $\lambda$4552 line. We note, however, that the relatively low signal-to-noise of the spectrum casts some doubt on the absence or on the detection of these two lines. This is particularly true for the Si {\small III} $\lambda$4552 which is placed between the relatively intense emission lines of Fe {\small I} $\lambda$4549 and Fe {\small I} $\lambda$4556. The presence of a strong Mg {\small II} $\lambda$4481 line would also hint at a type later than B2V. However, the presence of C {\small II} $\lambda$4267 indicates a B1.5V-B2V type and that of O {\small II} $\lambda$4070 line excludes types later than B2. In addition, the intense C {\small III} $\lambda$4650 and O {\small II} $\lambda$4640 absorption lines argue in favor of a spectral type hotter than B1V. A luminosity class V is consistent with the weakness of Si {\small IV} $\lambda$4089 line. To summarize, we constrain the spectral type to be between B0.5 and B2.5 with a most probable B1.5Ve type. 

Other conspicuous features in the spectrum of \object{HD\,157832} are the intense Fe and H emission lines (especially H$\alpha$, with an equivalent width (EW) $\sim$ 25\AA) which reveal the high density and large extent of the circumstellar disc. 
The equatorial plane of the circumstellar disc, and therefore also that of the star, are likely seen at intermediate-to-high inclination, as evidenced by the double peaked profiles seen in several emission lines (e.g., Fe {\small II} $\lambda$4233, 4584, 5018; and H$\alpha$; see Fig. \ref{fig1}). Using the FWHM of He {\small I} lines and the scaling relations derived by  \citet{Steele99}, we obtain {\it vsini} $\sim$ 266 km s$^{-1}$ from He {\small I} $\lambda$4471, and {\it vsini} $\sim$ 217 km s$^{-1}$ from He {\small I} $\lambda$4026, 4143, 4387. 

We built an overall spectral energy distribution using IUE LWP and SWP spectra acquired on September 12 1995 and Tycho, 2MASS (1.2-2.2$\mu$), AKARI (9-18 $\mu$) and IRAS (12-60 $\mu$) photometry. Fitting the 2200\AA\ bump together with the B \& V band photometry gives E(B-V) = 0.24 $\pm$ 0.01 \citep[in agreement with the value of 0.25 obtained by][]{Kozok85} and $T_{eff}$ = 25,000 $\pm$ 1,000 K. To that end, we used the reddening relation of \citet{Cardelli1989} and model atmospheres of \citet{Castelli2004}. The range of effective temperatures points at the same B1.5Ve spectral type as derived from our optical spectrum \citep{Bessel1998}. Redwards of 1.2$\mu$, the contribution of the circumstellar disc is clearly detected as a flux excess above stellar photospheric emission. 

Assuming a B1.5Ve spectral type, $M_V$ = -2.8 mag \citep{Humphreys84}, and E(B-V) = 0.24, we estimate a distance of 530 pc for \object{HD\,157832}.

\begin{figure}
\includegraphics[angle=-90,scale=.34]{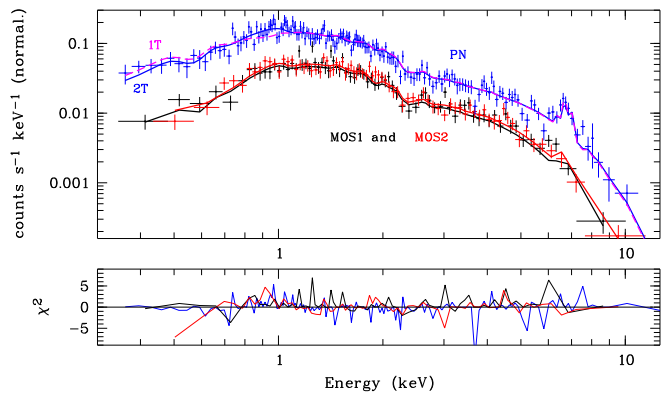}
\caption{EPIC/XMM spectra of HD\,157832. Solid lines represent the best 2-T model fit, while the dashed line corresponds to the 1-T model (only shown for the PN spectrum).\label{fig:spct}}
\end{figure}

\subsection{X-ray spectral analysis}

Spectral analysis was carried out with {\em XSPEC} v12.6.0. The EPIC spectrum displays a strong FeK$\alpha$ complex in emission with an intensity consistent with that expected from a hot thin thermal emission (see Fig. \ref{fig:spct}). 
Rough estimates of the equivalent widths are: EW(FeK) $\sim$ 47\,eV, EW(Fe {\small XXV}) $\sim$ 102\,eV, and EW(Fe {\small XXVI}) $\sim$ 36\,eV. The intensity of the ionized Fe {\small XXV} (6.67\,keV) and Fe {\small XXVI} (6.97\,keV) lines and continuum, especially the hard continuum, are well described by a simple 1-T ({\it mekal}) model multiplied by a photoelectric absorption component ({\it phabs}):
$kT$ = 10.01$^{+1.42}_{-1.17}$\,keV and $N_{H}$ = 2.1\,$\pm$\,0.1$\times$10$^{21}$\,cm$^{-2}$, 
with $\chi^2_{\nu}$/d.o.f. = 1.04/321. 
Abundances are consistent with solar values. However, the 1-T model under-predicts the emission observed in the $\sim$ 0.8--1.2\,keV range which contains several FeL-shell emission lines, and slightly over-predicts the soft E $<$ 0.7\,keV emission (see Fig. \ref{fig:spct}). The fit in both regions is clearly improved by including a second thermal component. Although the decrease of the $\chi^2_{red}$ from 1.04 to 0.94 is modest, it is statistically significant (F-test null hypothesis probability of 3.3$\times$10$^{-8}$). Both plasmas have abundances consistent with solar values: there is no evidence of a sub-solar Fe abundance similar to that observed in $\gamma$-Cas \citep{Smith04,Lopes10}.
A Gaussian line with $\sigma$ = 10\,eV and centered at 6.4\,keV was added to account for the tentative Fe K fluorescence line. The X-ray column density is about twice that predicted from the E(B-V) excess, indicating the presence of local photoelectric absorption. 

\begin{table}
\begin{center}
\caption{X-ray spectral parameters \label{tbl:parameters}}
\begin{tabular}{lcc}
\tableline\tableline
Parameter & 2-thermal\\
\tableline
$N_H$ (10$^{21}$\,cm$^{-2}$)                     & 2.3$\pm$0.2\\
$kT_{soft}$                                                   & 0.82$^{+0.16}_{-0.11}$\\
$f_x$ (10$^{-13}$\,erg\,cm$^{-2}$\,s$^{-1}$)  & 1.2\\
EM (10$^{55}$\,cm$^{-3}$)                           &  0.04\\
$kT_{hot}$                                                   & 11.25$^{+2.40}_{-1.57}$\\
$f_x$ (10$^{-12}$\,erg\,cm$^{-2}$\,s$^{-1}$)  & 3.7\\
EM (10$^{55}$\,cm$^{-3}$)                           & 1.1\\
$\chi^2_{\nu}$/d.o.f.                                              & 0.94/319\\
\tableline
\end{tabular}
\end{center}
\end{table}

Similarly to other $\gamma$-Cas-like stars, the X-ray spectrum of \object{HD\,157832} is dominated by the hot thin thermal component which accounts for $\sim$ 97\% of the total flux and emission measure (Table \ref{tbl:parameters}). 
The total unabsorbed flux of 3.8$\times$10$^{-12}$\,erg\,cm$^{-2}$\,s$^{-1}$ (0.2--12\,keV) corresponds to a luminosity of $\sim$ 1.3$\times$10$^{32}$\,erg\,s$^{-1}$ at 530\,pc.

\subsection{Timing analysis}

\begin{figure}
\includegraphics[angle=-90,scale=.60]{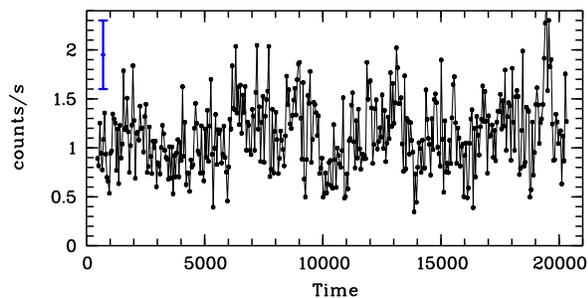}
\caption{EPIC (PN+MOS1+MOS2) light curve at 0.3--10\,keV. Time bin of 50\,s. A typical error bar is shown at the top-left.\label{fig:lc}}
\end{figure}

Timing analysis was performed on both PN eventlist and binned (10\,s) light curves created by summing all EPIC cameras in the energy bands 0.3--10\,keV, 0.3--2\,keV, and 2--10\,keV. In all cases, photon arrival times were corrected to the barycenter of the solar system. The light curves were corrected for inefficiencies in the mirror-detector system and background subtracted with the \textsc{epiclccorr}/SAS task v1.4.7. 
We searched for short periodicities up to the Nyquist frequency of $\sim$ 6.81\,Hz using the $Z^{2}_{n}$ periodogram \citep{Buccheri83} applied to the PN eventlist. 
No significant signal was found in the range of periods from 0.147\,s to 20\,s with a pulsed fraction larger than 14\% (0.3-10\,keV).
Binned light curves were mainly used to investigate hardness ratio variations with time and intensity, and to search for ``long-term'' periodicities ($P$ $>$ 20\,s). A wealth of peaks are seen in the Scargle periodograms \citep{Scargle82} with unfortunately only low statistical probabilities to be associated with a true period. 
Random variations are clearly seen in the X-ray light curve (see Fig. \ref{fig:lc}). The intensity can suffer an increase (decrease) of $\sim$ 140\% ($\sim$ 80\%) on time scale as short as 50 s.
Hardness ratios do not show correlations with the intensity of the source.

\section{Long term X-ray variations}

\object{HD\,157832} was serendipitously observed in the XMM-{\it Newton} slew survey \citep[XMMSL1;][]{Saxton08} on 2004 February 26. The 0.2--12\,keV flux of 3.4$\pm$1.0 $\times$10$^{12}$\,erg\,s$^{-1}$\,cm$^{-2}$ detected at this occasion is consistent with that measured in 2008. Most interestingly, the star was not detected in the ROSAT all-sky survey. Scanning observations took place in August/September 1990 for a total of $\sim$ 400\,s (derived from the exposure maps) spread over a time interval $\ga$ 2\,days. A faint low likelihood source, RXS J172754.1-470246, is present at only 1.2\arcmin\ from \object{HD\,157832} and is not detected again in the XMM-{\it Newton} EPIC images. Although RXS J172754.1-470246 is located too far to be associated with the new $\gamma$-Cas-like star the tail of its PSF adds some photons to the area enclosed by \object{HD\,157832}. We estimated a 90\% confidence upper limit of 0.02\,cts s$^{-1}$ in the 0.1-2.4\,keV ROSAT PSPC band for \object{HD\,157832} using the {\em uplimit} task \citep{Gehrels1986} in the {\em XIMAGE}\footnote{http://heasarc.gsfc.nasa.gov/xanadu/ximage/ximage.html} package. Using WEBPimms v4.2a we find that if the star would have had the X-ray flux seen by XMM-{\it Newton} at the time of the ROSAT all-sky survey, it would have been detected with a count rate of 0.055 cts s$^{-1}$. \object{HD\,157832} was thus on average at least 2.7 times fainter in 1990 than in 2008 and in 2004.

\section{Discussion}
\label{sct:discussion}

$\gamma$-Cas-like stars display astonishingly homogeneous X-ray and optical properties \citep{M07}. \object{HD\,157832} clearly exhibits all key characteristics of this class of objects, namely, an unusually hard thermal X-ray spectrum with a luminosity in the range of $\sim$ 10$^{32-33}$\,erg\,s$^{-1}$ variable on short time scales, and on the optical side, an early B type spectrum with a dense circumstellar disc. \object{HD\,157832} is thus a new $\gamma$-Cas-like Be star. 

However, ignoring the two faintest objects \citep[\object{USNO\,0750-13549725} and \object{SS\,397};][]{Lopes06} for which only low resolution optical spectra and uncertain classification are available, all other $\gamma$-Cas analogs, \object{HD\,161103}, \object{SAO\,49725}, \object{HD\,110432} \citep{SB06}, \object{HD\,119682} \citep{Safi07}, and \object{$\gamma$-Cas} itself have a B0.5e spectral type with a luminosity class III to V. \object{HD\,157832} is therefore the ``coolest'' $\gamma$-Cas analog identified so far. 

The lack of detection in the ROSAT all-sky survey over a $\ga$ 2\,day time interval shows that the soft (0.1-2.4\,keV) X-ray luminosity of \object{HD\,157832} varied quite significantly on a time scale of several years. Variability by a factor of $\sim$ 3 has been observed from \object{$\gamma$-Cas} itself in the hard (2-30\,keV) band of RXTE on a time scale of 50 to 90 days \citep{SHV06}. These modulations were accompanied by cycles in the B and V optical bands strongly suggesting that X-rays and optical variations have intimately related mechanisms. ROSAT pointed observations of \object{$\gamma$-Cas} revealed a factor 3 variability of the (0.1-2.4\,keV) band flux on a time scale of a few hours. However, the average PSPC count rate changed by only a factor $\sim$ 1.5 between all-sky survey and pointed observations \citep{Haberl1995}. Therefore, our lower limit of 2.7 on the soft band long term variability of \object{HD\,157832} is in principle compatible with what is observed from \object{$\gamma$-Cas}, but could also indicate a much larger variation of the soft X-ray flux. 

We did not find any information on the status of the circumstellar disc (CD) of \object{HD\,157832} close to the time of the ROSAT observations. However, both the IRAS and 2MASS measurements obtained in 1983 and 1999, respectively, show an excess of infrared emission above the stellar continuum and thus indicate the presence of a dense CD. The 16\,yr time interval between the two infrared observations is long enough to accommodate a normal B phase such as the one undergone by \object{$\gamma$-Cas} from 1942 till 1946 \citep{Doazan83}.  Alternatively, a factor 7 increase in the photo-electric absorption is enough to decrease the ROSAT PSPC count rates to 0.02 cts s$^{-1}$ while keeping the normalization of the X-ray spectrum to its 2008 value. \cite{Lopes10} report an even stronger change in the column density affecting at least 25\% of the hot X-ray component of \object{$\gamma$-Cas}.

The discovery of a $\gamma$-Cas-like star displaying long term X-ray flux variations with an amplitude comparable to and maybe higher than that of $\gamma$-Cas itself opens the possibility to study in depth how X-rays correlate with emission at other wavelengths. For $\gamma$-Cas itself, X-ray, UV, and optical fluxes were correlated one with each other, but
not with the orbital phase \citep{RSH02}. This behavior, together with the presence of migrating subfeatures in the line profiles of $\gamma$-Cas and \object{HD\,110432} \citep{SB06} give strong support to the magnetic scenario.

It is worth noticing that in all cases where hard X-ray emission was observed from $\gamma$-Cas-like stars, contemporaneous optical observations indicated the presence of a well developed circumstellar disc. 
Actually, the presence of the disc is a fundamental ingredient in both the ``magnetic'' and ``binary'' scenarios proposed to explain the X-ray emission of $\gamma$-Cas-like stars. While the difference in rotational velocity between the photosphere and the internal parts of the disc accounts for magnetic field winding, reconnection and particle acceleration in the magnetic model, the presence of a dense and extended CD is obviously also required in the accretion paradigm.

A multiwavelength campaign during disc dissipation, total disappearance, and disc rebuilding phases (Be $\rightarrow$ B $\rightarrow$ Be transition) will provide us with a definite answer about the role of the disc and very likely about the origin of X-ray emission in $\gamma$-Cas-like stars. 
For example, in the accreting scenario, the drop in X-ray luminosity associated to the disc-less episode should not be correlated with strong temperature variations of the hottest X-ray component -- since this property is mainly set by the gravitational potential of the compact star. In the magnetic scenario, we expect residual X-ray emission due to active sites, tentatively similar to that observed from some magnetic O-type stars \citep[e.g.,][]{Donati09}, and therefore hard enough to be distinguished from the soft emission usually observed in isolated massive stars.

The study of \object{HD\,157832} is relevant in at least two respects. First, the unique smaller effective temperature of \object{HD\,157832} compared to those of other $\gamma$-Cas analogs opens the possibility to study how the mass of the central star affects the X-ray emission properties and the evolutionary status of a putative accreting binary. Second, a monitoring of its remarkable variable and perhaps transient X-ray emission can shed light on the dependence of the X-ray emission on CD properties.
Therefore, \object{HD\,157832} seems to be a promising target for unveiling the true nature of $\gamma$-Cas-like stars.

\acknowledgments

R.L.O. acknowledges financial support from the Brazilian agency FAPESP 
	(\emph{Funda\c c\~ao de Amparo \`a Pesquisa do Estado de S\~ao Paulo}) 
	through a Young Investigator Program 
	(numbers 2009/06295-7 and
	2010/08341-3). 
We gratefully acknowledge the team of the Brazilian {\it Pico dos Dias} Observatory and National Astrophysics Laboratory for promptly observing our target in Director's Time and for kindly conducting the optical observations.	It is our pleasure to thank Myron A. Smith for very helpful discussions and comments.
	We also thank the XMM-{\it Newton} User Support Group, in particular Pedro Pascual, for the help with the SASv10.

{\it Facilities:} \facility{Simbad}, \facility{Aladin}, \facility{Vizier}.

\end{document}